\documentclass[aps,prb,twocolumn,nopacs,amssymb]{revtex4}

\usepackage[dvips]{graphicx}
\usepackage{dcolumn}
\usepackage{bm}
\def\be{\begin{equation}}
\def\en{\end{equation}}

\def\p{\partial} 
\newcommand{\av}[1]{\langle{#1}\rangle}

\def\gs{\gtrsim}
\def\ls{\lesssim}
\newcommand{\bi}[1]{\mbox{\boldmath$#1$}}
\def\p{\partial}
\def\bea{\begin{eqnarray}}
\def\ena{\end{eqnarray}}
\newcommand{\ppp}[3]{{\bigg(}\frac{\partial {#1}}{\partial {#2}}{\bigg )}_{#3}}

\renewcommand{\theequation}{\arabic{section}.\arabic{equation}}

\begin{document}
\title{Henry's  law,  surface tension, and surface adsorption 
 in   dilute binary mixtures}  
\author{Akira Onuki}
\email[]{onuki@scphys.kyoto-u.ac.jp}
\affiliation{Department of Physics, Kyoto University, Kyoto 606-8502,
Japan}



\date{\today}

\begin{abstract}

 Equilibrium  properties 
of  dilute binary fluid mixtures  are studied 
in two-phase states   
on the basis of  a   Helmholtz free energy 
 including the gradient free energy. 
The solute  partitioning between 
 gas and liquid  (Henry's law) and  
the surface tension change $\Delta\gamma$ 
are discussed. A derivation of  
the  Gibbs law 
 $\Delta\gamma=-T\Gamma$ is given with $\Gamma$ being  
the surface adsorption. 
Calculated quantities include 
the derivatives $d T_c/dX$ and $d p_c/dX$ 
 of the critical temperature and 
pressure with respect to the solute molar fraction $X$ 
and  the temperature-derivative  $(\p \gamma/\p T)_{{\rm cx},p}$   
of the surface tension  
 at fixed pressure $p$ on the 
coexistence surface.
Here   $(\p \gamma/\p T)_{{\rm cx},p}$      
can be both positive and negative, 
depending on  the solute molecular size 
and the solute-solvent  interaction, 
  and diverges on the azeptropic line. 
Near the solvent critical point,  
it is proportional to $(dp_c/dX)/K_{\rm Kr}$, where 
 $K_{\rm Kr}$ is  the Krichevskii parameter. 
Explicit  expressions are given for all these quantities 
in the van der Waals model. 
\end{abstract}


\maketitle


\section{Introduction}
\setcounter{equation}{0}

Many problems in physics and  engineering involve 
dilute solutions.  
In one-phase states, the critical behavior of dilute  
fluid mixtures have been studied 
extensively \cite{Griffiths,Anisimov,Onukibook}, where 
crossover occurs from pure-fluid behavior  to 
binary-mixture behavior 
 on approaching the critical line. 
 In two-phase states, 
it has been of great interest 
how a  solute is partitioned between  
gas and liquid  and how 
it is adsorbed in or repelled from the interface region 
\cite{Sengers1,SengersReview,De,Harvey}.

In the dilute  limit, the solute-solute  interaction 
may be neglected for nonelectrolytes. 
Nevertheless, the two-phase behavior is still highly 
nontrivial,   depending  sensitively on 
the detail of the solute-solvent  
interaction. In particular, the surface tension change $\Delta\gamma$ 
due to a solute  is related to the 
excess solute adsorption \cite{Gibbs}. 
 To understand 
such effects, we will  present a simple Ginzburg-Landau theory 
of dilute mixtures including the gradient free energy  \cite{Onukibook}. 
As is well-known \cite{vander}, van der Waals 
originally constructed such a theory 
for  pure fluids to describe 
a gas-liquid  interface and to calculate  
the surface tension $\gamma$.  For 
 binary mixtures it is moreover possible 
to calculate the solute density profile around an  
interface, which should satisfy  
 the Gibbs  adsorption law. In this approach 
solute partitioning between the two phases 
may be  examined  systematically.

In fluid hydrodynamics 
involving a gas-liquid interface, 
it is crucial how the surface tension 
varies on the surface as a function 
of ambient  temperature, 
 concentration, and  pressure,   
 since its variation induces  a 
Marangoni flow \cite{Levich,Straub}. 
However, the present author  is not aware 
of any fundamental theory   
on the surface variation of $\gamma$ in fluid mixtures 
in nonequilibrium. 
Hence we will also calculate the surface-tension derivative 
$(\p \gamma/\p T)_{{\rm cx}, p}$ with respect to 
the temperature  $T$ at fixed pressure  $p$ 
 in two-phase  coexistence \cite{Maran}.

In Section II, we will present a  Ginzburg-Landau model  
to calculate how the coexistence surface and the critical line 
are formed with addition of the second component. 
Mean-field critical behavior of dilute mixtures will 
be discussed, where the so-called Krichevskii parameter 
\cite{Sengers1,SengersReview,De,Harvey,Kri,OC,Shock} 
will be of  crucial relevance. On the basis of 
 the Gibbs adsorption 
law to be derived in Appendix A, 
general expressions for the surface tension 
variations on the coexistence 
surface will  be given.   In Section III, 
use will be made of 
the van der Waals free energy  of dilute mixtures 
\cite{Sengers,De} 
supplemented with the gradient free energy. 
It will give   explicit expressions for  
all the physical quantities  discussed 
in Section II, in terms of 
two dimensionless parameters characterizing 
the solute-solvent interaction. 
In Appendix B, correlation-function  expressions 
for  thermodynamic derivatives 
including that of the Krichevskii parameter 
will be given  \cite{Onukibook,OnukiJLTP}.

\section{Theoretical background}
\subsection{Ginzburg-Landau theory }
\setcounter{equation}{0}

This paper treats  dilute nonelectrolyte 
binary mixtures  with 
short-range interactions undergoing the gas-liquid transition. 
The number densities 
of the two components are written as $n_1$ and $n_2$ 
with $n_2\ll n_1$, 
which  are coarse-grained  variables  changing smoothly   in space. 
A  Ginzburg-Landau free theory is used to 
describe two-phase coexistence. 
A number of authors calculated the surface tension 
of mixtures by combining an equation of state and 
the gradient theory \cite{Carley,Sahimi,Strenby}.

Hereafter the Boltzmann constant will be set 
equal to unity. 
The free energy functional 
$F=F\{n_1,n_2\}$ depends on $n_1$ and $n_2$  as  
\be 
F= \int d{\bi r}\bigg[f+ \frac{T}{2}\sum_{i,j=1,2} D_{ij}\nabla n_i 
\cdot \nabla n_j \bigg]. 
\en 
The first term 
$f=f(n_1,n_2,T)$ in the brackets is the Helmholtz free 
energy density dependent on the densities and the 
temperature $T$. The gradient terms are 
needed to account for a free-energy increase 
due to density inhomogeneity.  
The coefficients  $D_{11}$, $D_{12}=D_{21}$, 
 and $D_{22}$ are assumed to be constants 
independent of the densities. 
In the dilute case $n_2\ll n_1$, 
the following form is assumed:   
\be
f= f_0(n_1,T)+ T n_2[\ln(n_2\lambda_2^3)-1 
+ \varphi(n_1,T)] .
\en 
Here   the van der Waals attractive 
interactions among the 
 molecules of the species 2 $(\propto n_2^2$) are neglected.  
The  $f_0(n_1,T)$ is the Helmholtz free energy density 
of the one-component (pure) fluid 
of  the species 1 and $\lambda_2= \hbar(2\pi/m_2T)^{1/2}$ 
($\hbar$ being the Planck constant) is the de Broglie 
length of the species 2. The term $Tn_2\varphi$ 
arises from the solute-solvent interaction, where 
$\varphi=\varphi(n_1,T)$ 
is independent of $n_2$ (see the next section 
for its van der Waals  expression).

For the free energy density $f$ in Eq.(2.2) 
the chemical potentials of the two components 
(without the gradient contributions)  are expressed as  
\bea 
\mu_1&=&\frac{\p f}{\p n_1}= 
 \mu_0 (n_1,T) + Tn_2 \varphi' (n_1,T), 
\\
{\mu}_2&=&\frac{\p f}{\p n_2}= 
 T\ln(n_2\lambda_2^3)+ T \varphi(n_1,T),
\ena 
where $\mu_0= \p f_0/\p n_1$ is the chemical potential of the 
pure fluid and 
$\varphi'=\p \varphi/\p n_1$ in $\mu_1$. Note that 
$\mu_2$ tends to $-\infty$ logarithmically 
in the low density limit 
$n_2\rightarrow 0$.  The pressure 
$p=n_1\mu_1+n_2\mu_2-f$ is expressed as 
\be 
p= n_1\mu_0 -f_0 + Tn_2 (1+ n_1\varphi') ,
\en 
where the last term is the solute correction. 
For the free energy functional $F$ in Eq.(2.1) 
the generalized chemical potentials including the gradient 
contributions read 
\be 
\hat{\mu}_i= \frac{\delta F}{\delta n_i}= 
\mu_i -T \sum_{j=1,2} D_{ij}\nabla^2 n_j \quad (i=1,2),  
\en 
which are  homogeneous 
in space in equilibrium.  
The usual chemical potentials $\mu_1$ and $\mu_2$ deviate from 
$\hat{\mu}_1$ and $\hat{\mu}_2$ in the interface region. 
Originally,  van der Waals 
set up the following interface 
equation  for pure fluids \cite{vander},
\be 
\mu_0(n,T)  - TD_{11}n'' = \mu_{\rm cx}^0(T),
\en  
where $ \mu_{\rm cx}^0$ is the chemical potential 
on the coexistence curve $p=p_{\rm cx}^0(T)$ of the pure fluid. 
The density $n(z)$ changes along the $z$ axis and $n''= d^2n/dz^2$.
Our equations in Eq.(2.6) lead to the van der Waals 
interface equation (2.7) for $n_1=n(z)$ and $n_2=0$.

In this paper  a small parameter $\zeta$ is defined as  
\be 
\zeta= \lambda_2^{-3} e^{\hat{\mu}_2/T}, 
\en 
which has the dimension of density. 
The solute density $n_2$ is expressed as  as 
\bea 
n_2 &=&  \zeta \exp\bigg[-\varphi(n_1,T)+ 
{D_{12}}\nabla^2n_1+ {D_{22}}\nabla^2n_2\bigg] 
\nonumber\\
&\cong &  \zeta \exp\bigg[-\varphi(n_1,T)+ 
{D_{12}}\nabla^2n_1\bigg].
\ena 
The fugacity of solute 
$f_2= \exp({\hat{\mu}_2/T})=\lambda^3_2\zeta$ is usually used to 
represent the degree of solute doping. 
The term proportional to 
$\nabla^2 n_2$ in the first line 
is  omitted in the second line. In the second line 
 $n_2$ is  expressed 
in terms of $n_1$  and $\nabla^2n_1$. 
 It follows  $n_2= \zeta \exp[-\varphi(n_1,T)]$ 
in the homogeneous bulk region.  
In our theory  
expansions up to first order in $\zeta$ or $f_2$ 
are performed.    On the other hand,  
 Leung and Griffiths \cite{Griffiths} 
used  another  parameter 
$\zeta_{\rm LG}\equiv  1/[1+A_0 \exp(\mu_1/T-\mu_2/T)]$ in order  to describe 
the overall thermodynamics of binary mixtures along the 
critical line ($0\le X\le 1$), where $A_0$ is  an appropriate 
 constant.

Equilibrium  states may be characterized 
in terms of  the field variables, 
$T$ and $\zeta$,   (instead of $T$ and the 
average solute density). 
As a functional of $n_1$ parameterized by $T$ and $\zeta$, 
 the grand  potential is defined as   
\be
\Omega =  F-\int d{\bi r}(\hat{\mu}_1n_1+\hat{\mu}_2n_2)  .
\en
In the dilute case  $\hat{\mu}_2$ 
 may be removed with the aid of 
Eqs.(2.4) and (2.6), leading to   
\be 
{\Omega} 
=  \int d{\bi r}\bigg [{f_0-\hat{\mu}_1n_1} +
\frac{T}{2}D_{11} |\nabla n_1|^2 
- Tn_2 \bigg],
\en  
where the gradient terms  
 proportional to $D_{12}$ 
cancel to vanish and  $n_2$ 
depends on $n_1$ as in the second line of Eq.(2.9). 
Here  $\Omega=\Omega(\{n_1\},T,\zeta)$ 
is minimized in equilibrium as a functional of $n_1$
In fact ${\delta \Omega}/{\delta  n_1}=0$ 
  holds from $\delta n_2= -n_2 
[\varphi'- D_{12}\nabla^2]\delta n_1$ 
at fixed $T$ and $\zeta$.

\subsection{Two-phase coexistence}

Let  a planar  interface 
separate   gas and  liquid 
regions. The bulk densities of the two components 
far from the interface are written as 
$n_{1\ell}$, $n_{1g}$, $n_{2\ell}$, and 
$n_{2g}$. The subscripts $\ell$ and $g$ stand for liquid 
and gas, respectively.  This paper treats  the dilute regime,  
\be  
n_{2 g}\ll n_{1 g},\quad 
n_{2\ell}\ll n_{1\ell},
\en 
in the two phases. 
Hereafter  thermodynamic relations 
in this case are given. For a noncondensable gas as a solute, 
another typical  situation  is 
given by $ n_{1 g} \ls n_{2 g} \ll n_{1\ell}$  
far below the solvent criticality.

As a reference state, we consider 
 the two-phase state of  the  pure fluid 
composed of the first component  
at the same temperature $T$ below $T_{c0}$,  
where  $n_1$ is equal to $n_{\ell 0}$ in liquid 
and $n_{g 0}$ in gas.  
The chemical potential and pressure  
in the pure fluid are written as $\mu_{\rm cx }^0$ and 
$p_{\rm cx}^0 $, respectively. 
With addition of solute, 
Eq.(2.9) yields the bulk solute densities, 
\be 
n_{2\ell}= \zeta e^{-\varphi_{\ell}}, 
\quad n_{2g}=
\zeta e^{-\varphi_g}.
\en  
where $\varphi_\alpha= 
\varphi (n_{\alpha 0},T)$ with 
 $\alpha$ standing for $\ell$ or $g$.  
Since the pressure is given by the common value 
$p=p_{\rm cx}$ in the two phases, 
Eq.(2.5) yields the coexisting solvent densities 
$n_{1\alpha}$   ($\alpha=\ell$ or $g$)  as   
\be 
\frac{n_{1\alpha}}{n_{\alpha 0}}-1=K_{T\alpha}\bigg
[\delta p_{\rm cx}- 
T n_{2\alpha}(1+n_{\alpha 0}\varphi'(n_{\alpha 0}))\bigg],
\en 
where   $K_{T\alpha}$ is the isothermal compressibility 
$K_T=(\p n/\p p)_T/n$ 
of the pure fluid  for  $\alpha=\ell$ or $g$ and 
$\delta p_{\rm cx}=p_{\rm cx}-p_{\rm cx}^0$ 
is the deviation of the coexisting pressure. 
Furthermore, Eq.(2.3) yields 
\be 
\delta\mu_{1{\rm cx}}=\mu_{1{\rm cx}}-\mu_{{\rm cx}}^0=
 (\delta p_{\rm cx}
-Tn_{2\alpha})/n_{\alpha 0}
\en 
for the deviation of the solvent chemical potential 
in two-phase coexistence. This holds both for $\alpha=\ell$ and 
$g$, so 
$\delta p_{\rm cx}(1/n_{\ell 0}-1/n_{g0})-T(n_{2\ell}/n_{\ell 0}
-n_{2g}/n_{g 0})=0$. Thus,    
\bea 
&&\delta p_{\rm cx} = {T}{\Delta X}/{{\Delta v} },\\ 
&&\delta \mu_{1\rm cx}= {T}{\Delta n_2}/{{\Delta n} },
\ena 
where   $\Delta n= n_{\ell 0}-n_{g 0}$ 
and $\Delta v=1/n_{g 0} - 1/n_{\ell 0}$ 
are the differences  of the  density and the 
 volume (per particle)  between gas and liquid 
in  the pure fluid, respectively, (taken to be positive). 
The  differences 
of the solute density and molar fraction are written as 
\bea
\Delta n_2 &=& n_{2g}-n_{2\ell}= (e^{-\varphi_g}-
e^{-\varphi_\ell})\zeta,\\ 
\Delta X&=& \frac{n_{2g}}{n_{g 0}}-\frac{n_{2\ell}}{n_{\ell 0}} 
=\bigg(\frac{e^{-\varphi_g}}{n_{g 0}}-
\frac{e^{-\varphi_\ell}}{n_{\ell 0}}\bigg)\zeta,
\ena
which are both proportional to $\zeta$ from Eq.(2.13).

For infinitesimal variations of $\mu_1$, $\mu_2$, $T$, 
and $p$,  the Gibbs-Duhem  relation  generally holds in the form,  
\be 
d\mu_1=  -Xd\Delta -s dT + v dp,
\en  
where $\Delta=\mu_2-\mu_1$ is the chemical potential difference, 
$s$ is the entropy per particle,  and $v=1/(n_1+n_2)$ 
 is the volume per particle.  
In particular, for variations on  the coexistence surface 
in the $p$-$T$-$\Delta$  space, 
 we obtain 
\be 
\Delta X d\Delta = -\Delta s dT + \Delta v dp,
\en 
where $\Delta s$ and   $\Delta v$   
may be taken as 
the entropy difference of  the pure fluid. 
Here $d\Delta \cong T\zeta^{-1}d\zeta$ 
in the dilute case, so in the mixture case $\zeta>0$ we have 
\be
\ppp{T}{\zeta}{{\rm cx},p}= - T\frac{\Delta X}{\zeta\Delta s}, 
\quad  
\ppp{p}{\zeta}{{\rm cx},T}=  T\frac{\Delta X}{\zeta\Delta v}, 
\en  
where $(\p \cdots/\p \cdots)_{{\rm cx},p}$ 
and  $(\p \cdots/\p \cdots)_{{\rm cx},T}$ 
are the derivatives  on the coexistence surface 
at fixed $p$ and $T$, respectively, and 
the right hand sides of Eq.(2.22) are independent of $\zeta$ 
since $X\propto \zeta$.  Obviously, $\delta p_{\rm cx}$ in Eq.(2.16) 
follows from 
integration of $(\p p/\p \zeta)_{{\rm cx},T}$ in Eq.(2.22) with 
respect to $\zeta$ from the reference pure fluid state 
 at fixed $T$.
To derive $\delta \mu_{1{\rm cx}}$ in Eq.(2.17) we integrate  Eq.(2.20) 
with respect to $\zeta$ at fixed $T$ to obtain  
Eq.(2.15).  
Here note  the relation  
$\int_{-\infty}^{\Delta} X d\Delta 
\cong T \int_0^\zeta d\zeta X/\zeta= TX$, where 
$X/\zeta$ is independent of $\zeta$. 
In the same manner, 
the temperature change $\delta T_{\rm cx}
= T_{cx}(p,\zeta) -T_{cx}^0(p)$ 
at fixed $p$ (below the critical pressure $p_{c0}$) 
on the coexistence surface reads  
\be 
\delta T_{\rm cx}= 
-T {\Delta X}/{\Delta s},
\en 
which is proportional to $\zeta$.

It is convenient to introduce 
the  partition coefficient 
of solute  ${\cal K}$ as 
the ratio of the solute molar fraction 
in gas $X_g= {n_{2g}}/{n_{g 0}}$ and 
that in liquid $X_\ell={n_{2\ell}}/{n_{\ell 0}}$ \cite{Sengers1,SengersReview}.
 Equation (2.13) gives  
\be 
{\cal K}= \frac{X_{g}}{X_{\ell}}
=\frac{n_{\ell 0}}{n_{g 0}} \exp(\varphi_{\ell}- 
\varphi_g).
\en 
Then $\Delta X= (1- {\cal K}^{-1})X_g
=({\cal K}-1)X_\ell$. 
The azeotropic line on the coexistence surface is determined 
by ${\cal K}=1$, on which the two phases have the same composition. 
If the gas region is dilute, $X_g$ 
is nearly equal to the partial pressure 
of the second component divided by the total pressure in 
the gas region. Near the critical point ${\cal K}\to 1.$ 
When the gas phase is dilute,  
Henry's  constant $\cal H$ is usually 
defined as 
\be 
 {\cal H}=p_{2g}/ X_{\ell},
\en 
with  $p_{2g} (\cong 
Tn_{2g})$ being  the partial pressure of the solute, 
Here   ${\cal H}= p_g {\cal K}$, where 
$p_g$ is the total gas pressure.  
To analyze data near the critical point 
Sengers {\it et al.}\cite{Sengers1} used 
another definition of Henry's  constant,  
\be 
k_H= f_2/X_\ell,
\en 
where $f_2$ is the solute fugacity. 
In our notation we obtain 
$k_H = {\cal K}n_{g0}\lambda_2^3 e^{\varphi_g}$ 
from Eqs.(2.8) and (2.9).

\subsection{Surface tension and surface adsorption}

The  surface tension $\gamma$ 
in binary mixtures may be calculated from  Eq.(2.1). 
It has been calculated in  the gradient theory in 
 fair agreement with 
experimental data over a wide temperature range 
 \cite{Carley,Sahimi,Strenby,Kie}. 
However, our result cannot be used 
in the asymptotic critical 
region.

In Appendix A,  the deviation 
$\Delta\gamma=\gamma-\gamma_0$ will be calculated, 
where $\gamma_0$ is the surface 
tension in the pure fluid. 
For small $\zeta$ it follows 
the Gibbs relation \cite{Gibbs}, 
\be 
\Delta\gamma= -T\Gamma.
\en 
Here $\Gamma$ is the excess adsorption of the solute  
on the interface expressed as   
\be 
\Gamma=  
\int dz\bigg[ n_2(z) -n_{2\ell}+ \frac{\Delta n_2}{\Delta n}(n(z)-n_{\ell 0})
\bigg],
\en 
where $\Delta n_2=n_{2 g}-n_{2 \ell}$ and  
$\Delta n=n_{\ell 0}-n_{g 0}$ and the 
 integrand is nonvanishing far from the interface.

The physical meaning of $\Gamma$ is as follows. 
For   a finite system with length $L$ 
much longer than the interface width, 
the interface position 
$z=z_{\rm in}$ may  be determined 
with the aid of the Gibbs construction, 
\be 
z_{\rm in}n_{\ell 0}+(L-z_{\rm in})n_{g 0}= 
\int_0^L dz n(z). 
\en 
Then $\Gamma$ is expressed as   
\be 
\Gamma 
=\int_{0}^{z_{\rm in}}
 dz[ n_2(z) -n_{2\ell}] +
\int_{z_{\rm in}}^L 
 dz[ n_2(z) -n_{2g}],
\en 
where the first (second) 
term represents  the excess adsorption in the liquid (gas) region. 
The integrands here tend 
to 0 far from the interface, so 
we may push  the lower bound in the first integral to $-\infty$ 
and the upper  bound in the second  integral to $\infty$ 
for  a macroscopic system. 
The Gibbs relation (2.27) has been used frequently 
for surfactants added  in water-air 
and water-oil systems  \cite{Safran},  which 
induce a dramatic decrease of $\gamma$  
even at extremely low bulk densities.  
If salt is added, 
$\Delta\gamma$   
contains an electrostatic contribution also 
\cite{OnukiJCP}.

The surface tension 
$\gamma=\gamma(T,\zeta)$ of mixtures 
is defined on the coexistence 
surface $p=p_{cx}(T,\zeta)$.  
Since $\Delta\gamma \propto \zeta$ in  Eq.(2.27), 
use of Eq.(2.22) gives the temperature derivative of 
$\gamma(T,\zeta)$ at fixed $p$ in the form, 
\be 
\ppp{\gamma}{ T}{{\rm cx},p}=
 \frac{d\gamma_0}{dT} 
+\frac{\Delta s }{\Delta X} \Gamma .
\en 
It is important that 
the second term is independent 
of $\zeta$ as well as the first term.  In  the azeotropic 
case $\Delta X=0$,  the second term 
in the right hand side tends to 
$\pm \infty$.

\subsection{Mean-field critical behavior}

\subsubsection{Landau expansion}

The mean-field critical behavior 
of dilute binary mixtures 
will then be examined near  
the critical point of 
the pure  fluid (solvent criticality). 
The  critical temperature, 
pressure,  and density at the solvent criticality 
are written as $T_{c0}$, $p_{c0}$, 
 and $n_{c0}$, respectively, in the pure fluid.  
The order parameter  is 
the solvent density deviation,  
\be 
\psi=n_1-n_{c0}.
\en 
Here  $T-T_{c0}$ and $\psi$ are assumed to be small. 
The Landau expansion of $f_0(n_1,T)$ is of the form, 
\be 
f_0= f_{0c}(T) + \mu_{0c}(T)  \psi + \frac{A_0}{2}(T-T_{c0}) 
\psi^2+\frac{B_0}{4} \psi^4 ,
\en
where $f_{0c}(T)=f_0(n_{c0},T)$ 
and $\mu_{0c}(T)=\mu_0(n_{c0},T)$ are 
the free energy density and the chemical potential 
 at the critical density,  respectively. The Gibbs-Duhem relation 
for one-component fluids yields 
\be 
\mu_{0c}(T)=\mu_{c0}- (s_{c0}-n_{c0}^{-1}p'_{\rm cx} )(T-T_{c0}),
\en 
where $\mu_{c0}$ is the critical chemical 
potential, $s_{c0}$ is the critical entropy, 
and $p'_{\rm cx}=( \p p/\p T )_{\rm cx}$ 
is the derivative of $p$ with respect to $T$ along 
the coexistence line at the solvent criticality. 
Use has been made of the relation 
$(\p p/\p T)_n\cong p'_{\rm cx}$ near the 
solvent criticality \cite{Onukibook}.

A small amount of the second component 
is then added as a solute. 
Near the solvent 
criticality, we expand the solute density $n_2$ in Eq.(2.9) as 
\be 
\frac{n_2}{\zeta} 
= [C_0 + C_1 \psi+ \frac{C_2}{2}
 \psi^2+ \frac{C_3}{3}
 \psi^3 ](1- D_{12}\nabla^2\psi ].
\en    
Here  we may set   $T=T_{c0}$,  since 
 the term $-Tn_2$ is already a small 
perturbation  in the grand potential (2.11).  
The coefficients $C_0$, $C_1$, $C_2$, 
and $C_3$ are obtained from the expansion of $e^{-\varphi}$ as 
\bea 
C_0&=&e^{-\varphi_{c}},\quad  C_1= -\varphi_{c}'C_0,
\quad  C_2= (\varphi_{c}'^2- \varphi_c'')C_0,\nonumber\\
C_3&=& \frac{1}{2}({3} \varphi_c'\varphi_c''- 
\varphi_c'''-\varphi_c'^3)C_0, 
\ena
where $\varphi_{c}'$, $\varphi_{c}''$, and 
$\varphi_{c}'''$ are the derivatives  
$\p\varphi/\p n_1$, $\p^2\varphi/\p n_1^2$, 
and $\p^3\varphi/\p n_1^3$ at the solvent criticality.  
The  critical solute density and molar fraction read 
\be 
n_{2c}=\zeta C_0,\quad X_c= \zeta C_0/n_{c0}.
\en

Equilibrium is obtained by minimization of 
the grand potential  $\Omega$ in Eq.(2.11), 
which is the integral of  the density 
 $\hat{\omega}=  f_0-\hat{\mu}_1n_1- Tn_2$ plus the gradient term. 
Here $\hat{\mu}_1$ should be expressed 
in terms of the macroscopically 
given pressure $p$ (not treated as a fluctuating 
variable), temperature $T$, and $\zeta$.
From the expression for $\mu_1$ in Eq.(2.3)  
some calculations give 
\be 
\mu_1\cong \mu_{c0}+  \frac{p-p_{c0}}{n_{c0}}- s_{c0}(T-T_{c0})-
\frac{T_{c0}}{n_{c0}} n_{2c}, 
\en 
in the bulk regions. 
This  relation also follows from 
integration of the Gibbs-Duhem relation (2.20) for mixtures.  
Note that  $\hat{\mu}_1$ is equal to the right hand side of Eq.(2.38) 
in the whole space.  The Landau expansion 
of $\hat\omega=  f_0-\hat{\mu}_1n_1- Tn_2$ is   now of  the form, 
\be 
\hat{\omega}
=-p_0(T,\zeta)  -n_{c0}^{-1}h \psi +  
\frac{A_0}{2}(T-T_{c}
) \psi^2+\cdots ,
\en
where $p_0(T,\zeta)= p(n_{c0},T,\zeta)$  is the pressure  
in Eq.(2.5) at $n=n_{c0}$ and 
$h$ has the meaning of  
 the ordering field. Use  of 
Eqs.(2.34) and (2.38) gives   
\be 
 h= p-p_{c0} -p'_{\rm cx}(T-T_{c0}) - T_{c0}(C_0- n_{c0}C_1)\zeta, 
\en 
In the third term of Eq.(2.39) 
$T_c= T_{c0}+\Delta T_c$ 
is the critical temperature with the shift,
\be
\Delta T_c= \zeta T_{c0} C_2/A_0. 
\en
Since  $h= 0$ at the criticality 
$T=T_c$ and $p=p_c$,  
the critical pressure shift $\Delta p_c=p_c- p_{c0} $  is calculated as 
\be
\Delta p_c= p'_{\rm cx} \Delta T_c+
T_{c0}(C_0- n_{c0}C_1)\zeta.
\en
to first order in $\zeta$.  
Since $\Delta T_c$, $\Delta p_c$, and $X_c$ 
are all linear in $\zeta$, 
 the derivatives of $T_c$ and $p_c$  
along the critical line 
are  given by  ${d T_c}/{dX} ={\Delta T_c }/{X_c}$ and  
${d p_c}/{dX} ={\Delta p_c }/{X_c}$. 
The critical line is characterized by $X=X_c(\zeta)$ in 
 Eq.(2.37), leading to   
\bea 
\frac{d T_c}{dX} &=& 
 n_{c0}T_{c0}\frac{ C_2}{A_0C_0},
\\ 
\frac{d p_c}{dX} &=&  
p'_{\rm cx} \frac{d T_c}{dX}+
{n_{c0}T_{c0}}\bigg(1- n_{c0}\frac{C_1}{C_0}\bigg),
\ena

In addition, from the third order term $(\propto C_3$) 
in the expansion of $n_2$ in Eq.(2.35), there arises 
a small shift of the critical solvent density as 
\be 
 n_{1c} - n_{c0}=  \zeta T_{c0} C_3/B_0.
\en  
If  we expand $\hat{\omega} =  f_0-\hat{\mu}_1n_1- Tn_2$ 
up to the quartic term 
and rewrite it  in powers of  $n_1-n_{1c}$, 
the third order term  
 should vanish.  However, this critical density  shift 
 does not affect the shifts of 
$T_c$ and $p_c$ to first order in $\zeta$. Also the coefficient of the 
gradient  term  in $\Omega$ is changed from  $D_{11}$ to 
\be 
D_{11}'= D_{11} -\zeta  D_{12}C_1.
\en  
This correction is irrelevant 
in the dilute limit.

\subsubsection{Krichevskii parameter and 
concentration fluctuations}

In the literature 
\cite{Sengers1,SengersReview,De,Harvey,Kri,OC,Shock}, 
use has been made of the 
thermodynamic derivative $(\p p/\p X)_{n T}$ with 
 $n=n_1+n_2$ and  $X=n_2/n$ to analyze the critical behavior 
in dilute mixtures \cite{Kri}.  From Eq.(2.5) it 
 is equal to $ Tn_1 (1+n_1\varphi')-n_1^2f_0'' $ in our approximation. 
It is known to  tend   to  a well-defined  limit, 
called  the  Krichevskii parameter,   as $\zeta \to 0$ 
at  the solvent criticality.  In terms of $C_0$ and $C_1$ in Eq.(2.41), 
it  is  expressed as  
\be 
K_{\rm Kr}\equiv  
\ppp{p}{X}{n T}^c=  T_{c0} n_{c0} (1-n_{c0}C_1/C_0)
\en 
From Eqs.(2.43) and (2.44) 
it follows the well-known relation  \cite{Sengers1,Kri},
\be 
K_{\rm Kr}= \frac{dp_c}{dX} - p'_{\rm cx} \frac{dT_c}{dX}.
\en 
 From Eq.(2.35) the solute molar fraction 
behaves as 
$X=n_2/n_1= \zeta(C_0+C_1\psi)/n_{c0}-n_{2c} \psi/n_{c0}^2+\cdots $ 
at $T=T_{c0}$.  For small $T-T_{c0}$ and $\psi$ it is expressed as 
\be 
\frac{X}{ X_c}= 1+A_m(T-T_{c0})  - (K_{\rm Kr}/n_{c0}^2T_{c0})\psi+\cdots,
\en 
where $A_m$ is a constant. 
In two-phase coexistence  this equation yields  
\be 
{\Delta X}/{\Delta v} =  ({K_{\rm Kr}}/{T_{c0}}) X_c, 
\en 
From Eq.(2.22) this  is the near-critical expression of 
 $\zeta (\p p/\p \zeta)_{{\rm cx},T}/T_{c0}
=(\p p/\p \Delta)_{{\rm cx},T} $ in the dilute limit.

In Table 1,  
 we show  experimental data of $ T_{c0}^{-1} dT_c/dX$, 
$(n_{c0}T_{c0})^{-1} dp_c/dX$,  
$(n_{c0}T_{c0})^{-1}K_{\rm Kr}$, and $dp_c/dX/K_{\rm Kr}$  
 for dilute mixtures  near the solvent 
criticality, where the solvent is  
CO$_2$ \cite{Russia} or H$_2$O \cite{Toluene,Russia3}. 
For CO$_2$ we have $T_{c0}=304$K,   
$n_{c0}T_{c0}= 26.1$MPa, and  $(\p p/\p T)_{\rm cx}/n_{c0}=1.97$, 
while  for H$_2$O we have  
 $T_{c0}=647.01$K,   
$n_{c0}T_{c0}= 96.0$MPa, and  $(\p p/\p T)_{\rm cx}/n_{c0}=1.81$. 
Thus  $dT_c/dX$,  $dp_c/dX$, and  $K_{\rm Kr}$ can be 
both positive and negative depending 
on the specific details of the two components. 
These quantities are very small for H$_2$O-D$_2$O mixtures \cite{Russia3},
where the two component are very alike. 
If the solute is H$_2$O and the solvent is D$_2$O, their  signs are 
simply reversed with their absolute values nearly unchanged.

In  two-phase  coexistence with general  compositions, 
the present author introduced the parameter  
\cite{OnukiJLTP,Onukibook},
\be 
\epsilon_{az} \equiv 
n_c\frac{\Delta X}{\Delta n} 
= -\frac{1}{n_c}\ppp{p}{\Delta}{{\rm cx},T},
\en
where $n_c=n_{1c}+n_{2c}$ is the 
critical density and $\Delta n= n_\ell-n_g$.  
The critical line 
under consideration is that of the gas-liquid criticality 
for  $|\epsilon_{az}| \ls 1$ and is that of 
the consolute  criticality 
for  $|\epsilon_{az}| \gs 1$.  
In the dilute limit $X\to 0$,  we have 
$\epsilon_{az} \cong -(K_{\rm Kr}/n_{c0}{T_{c0}}) X$. 
For $^3$He-$^4$He mixtures \cite{Griffiths,OnukiJLTP}, the relation 
 $\epsilon_{az} \cong -\frac{1}{3} 
X(1-X)$ roughly holds along the critical line, 
  where $X$ is the $^3$He 
molar fraction. Thus  $K_{\rm Kr}/n_{c0}{T_{c0}}$ is 
$1/3$ with  $^3$He being a solute and 
is $-1/3$ with  $^4$He being a solute. 
Thus $^3$He-$^4$He mixtures 
are nearly azeotropic  at any $X$ 
(even away from the critical line). The resultant  
 crossover effects have been observed 
in near-critical  $^3$He-$^4$He mixtures 
in statics and dynamics \cite{Meyer}.

On approaching  the  critical point, 
the thermal fluctuation of $\psi$  is enhanced 
with its variance proportional to the  
compressibility $K_{T\Delta} =(\p n/\p p)_{T\Delta}/n$ 
as in Eq.(B5) in the appendix. 
As shown in  Eq.(2.49) or in Eq.(B12),  
the thermal fluctuation of 
the molar fraction contains the growing 
part $-(K_{\rm Kr}/n_{c0}^2T_{c0})X \psi$ 
\cite{Anisimov,Onukibook,OnukiJLTP}. 
From Eqs.(B5), (B8), and (B12) 
  the concentration susceptibility 
$(\p X/\p \Delta)_{pT}$ 
behaves  near the  criticality  as 
\be 
T\ppp{X}{\Delta}{pT}\cong 
X  + X^2 (K_{\rm cr}^2/n_{c0}T_{c0})K_{T\Delta}. 
\en 
The first term is the low density limit (see Eq.(B8)). 
The second is the singular contribution 
stemming   from the solute-solvent interaction. 
We may set 
$({\p X}/{\p\Delta})_{pT}\cong X$  and replace  
the mixture compressibility $K_{T\Delta}$ 
by the pure-fluid compressibility $K_{T}$  when 
\be 
 X K_{\rm cr}^2 K_T/n_{c0}T_{c0} \ll 1.
\en  
This condition  has been assumed in the 
definition 
of the Krichevskii parameter (see the appendix).  

\begin{table}
\caption{$T_c'/T_{c0}$, 
$p_c'/n_{c0}T_{c0}$, 
$K_{\rm Kr}/n_{c0}T_{c0}$, and  
$p_c'/K_{\rm Kr}$ for  CO$_2$+ solute 
and for H$_2$O + solute   
near the solvent critical point, where  
$T_c'= dT_c/dX$ and $p_c'= dp_c/dX$. 
The last quantity is related to the 
temperature-derivative of the surface tension 
in Eq.(2.60). 
Data are taken from Refs.\cite{Russia,Toluene,Russia3}. 
}
\begin{tabular}{cccccc}
\hline
Solvent&
 Solute & $T_c'/T_{c0}$ &
 $p_c'/n_{c0}T_{c0}$ & $K_{\rm Kr}/n_{c0}T_{c0}$
& $p_c'/K_{\rm Kr}$ \\ 
\hline 
  CO$_2$&Neon&$-0.0517$ &0.919 &1.02 & 0.900\\
\hline
 CO$_2$&Argon&$-0.192$
 & 0.553 & 0.936 & 0.591\\
\hline
CO$_2$&Ethanol  & 0.539
 & 0.694 & $-0.380$ & $-1.81$ \\
\hline
 CO$_2$& Pentanol & 2.20
 & 1.96 & $-2.42$& $-0.809$ \\
\hline
CO$_2$&  Ethane & $-0.182$
 & $-0.187$ & $0.175$& $-1.07$ \\
\hline
H$_2$O & Toluene & $-1.32$
 & $-0.948$ & $1.434$& $- 0.661$ \\
\hline
H$_2$O & D$_2$O & $-0.0050$
 & $-0.0041$ & $0.0050$& $-1.21$ \\
\hline
\end{tabular}
\end{table}

\subsubsection{Critical behavior of 
surface tension}

Using the Landau expansion of $f_0$ in Eq.(2.33) we 
 next examine the mean-field 
critical behavior in two-phase coexistence, 
where the average order parameter values 
in the two  phases are $\psi=\pm \psi_e$  with 
\be 
\psi_e= [A_0(T_{c0}-T)/B_0]^{1/2}.
\en  
The surface tension    of the pure fluid $\gamma_0$ is written as 
\be 
\gamma_0=\frac{4}{3}(T_{c0}-T)A_0 
\psi_e^2 \xi. 
\en 
The  interface profile is 
expressed as  
$\psi(z)= \psi_e \tanh (z/2\xi)$ along the surface normal, 
where  $\xi$ is 
the correlation length   in two-phase coexistence 
expressed as 
\be 
\xi= (D_{11}/2A_0)^{1/2} (1-T/T_{c0})^{-1/2}.
\en 
Thus $\gamma_0\propto 
(1-T/T_{c0})^{3/2}$, as originally derived by van der Waals \cite{vander}.

It is easy to calculate the surface 
adsorption $\Gamma$ in Eq.(2.28). 
Use of the expansion (2.35)  gives $\Gamma \cong 
\zeta C_2 \int dz [\psi(z)^2-\psi_e^2]$. Thus,  
\be 
\Gamma= -2C_2\psi_e^2\xi\zeta = -2\frac{\Delta T_c}{T_{c0}}A_0
 \psi_e^2\xi,
\en  
so $\Gamma \propto \zeta (1-T/T_{c0})^{1/2}$. 
Because 
$d \gamma_0/d T= -3\gamma_0/2(T_{c0}-T)= 
-2A_0 \psi_e^2\xi$ from Eq.(2.55), we find 
\be 
\Gamma =  \frac{d\gamma_0}{dT}\frac{\Delta T_c}{T_{c0}}.   
\en 
If we write $\gamma_0=A_s (1-T/T_{c0})^{3/2}$ 
with $A_s$ being a constant, 
the surface tension of 
dilute mixtures $\gamma=\gamma_0-T\Gamma$ is 
expressed as 
\be 
\gamma=
 A_s T_{c0}^{-3/2}[T_c(\zeta)-T]^{3/2},
\en 
to first order in $\zeta$. 
That is, the solute effect on $\gamma$ 
is only  to shift $T_{c0}$ to 
 $T_c(\zeta) =T_{c0}+\Delta T_c$.  
From Eqs.(2.31) and 
  (2.58) we may express 
$(\p \gamma/\p T)_{\rm cx}$ in terms of $\Delta T_c$. 
Further using Eq.(2.42) it assumes a simpler form in terms 
of $\Delta p_c$ or $d p_c/dX$ as 
\be 
\ppp{\gamma}{ T}{{\rm cx},p}\bigg/ \frac{d\gamma_0}{dT} 
= \frac{\Delta v}{\Delta X}
\frac{\Delta p_c}{T_{c0}}
 = \frac{1}{K_{\rm Kr}}\frac{d p_c}{dX}, 
\en 
which tends to a well-defined limit 
at the solvent criticality. 
See the last column of Table 1 for the above ratio. 
It is negative if $K_{\rm Kr}$ and 
${d p_c}/{dX}$ have different signs.

\section{van der Waals theory of mixtures}
\setcounter{equation}{0}

\subsection{Dilute mixtures}

The van der Waals theory of 
one-component fluids \cite{vander} 
was extended to binary mixtures 
by van der Waals and Korteweg \cite{Sengers,Onukibook}.  
For binary mixtures the Helmholtz free energy density 
$f={f}(n_1,n_2,T)$ is given by 
\be 
{f} = T \sum_i n_i \bigg [\ln \bigg(\frac{n_i\lambda_i^3}{1-\phi} 
\bigg )-1\bigg] -\sum_{ij} {w_{ij}} n_in_j,  
\en 
where $\lambda_i= (2\pi/m_iT)^{1/2}\hbar$ 
are the de Broglie lengthse with  $m_1$ and $m_2$ 
being the molecular masses and $\hbar$ being the Planck 
constant.   The $\phi=v_{10}n_1+v_{20}n_2$ is 
 the volume fraction of the 
hard-core region with  
$v_{10}$ and $v_{20}$  
being  the molecular volumes. 
The coefficients 
$w_{ij}$ represent 
the strength of the van der Waals attractive 
interaction between $ij$ pairs.  
However,  more elaborate thermodynamic models  
have  been used to predict the surface tension of 
real binary mixtures 
 \cite{Carley,Sahimi,Strenby}.

In the pure fluid limit ($n_2=0$), 
the free energy density and the chemical potentials are 
given by  
\bea 
f_0(n,T)& = & Tn \ln\bigg[\frac{n\lambda_1^3}{1-\phi}\bigg]
-Tn -w_{11}n^2,\\
\mu_{0} (n,T) &=& 
  T\ln\bigg[\frac{n\lambda_1^3}{1-\phi}\bigg] 
+\frac{T\phi}{1-\phi}- 
2w_{11} n,
\ena 
where we set $n=n_1$ and $\phi=v_{01}n_1$. 
Hereafter   
\be 
\epsilon= v_{10}^{-1} w_{11} 
\en 
is the attractive energy among the molecules of the first 
component. In the pure fluid,  the critical 
temperature,  pressure, and density 
are written as 
\be 
T_{c0}= \frac{8\epsilon}{27},
\quad p_{c0}= \frac{\epsilon}{27}v_{10}^{-1}, 
\quad n_{c0}= \frac{1}{3}v_{10}^{-1}.
\en  
See the upper plate of Fig.1 for the liquid and gas densities 
in the van der Waals model. 
Far below the critical temperature in two-phase coexistence, 
 the gas density $n_{g0}$ becomes 
very small compared to the liquid density 
$n_{\ell 0}$. 
In fact, if $p_{cx}^0 \cong Tn_{g0} \ll Tn_{\ell 0}$. 
 the van der Waals theory yields  
\bea 
&&\phi_\ell \cong \frac{1}{2} +\frac{1}{2} (1-4T/\epsilon)^{1/2}, \\
&&\phi_{g}/\phi_{\ell}  \cong 
(\epsilon \phi_\ell/T)e^{-\epsilon \phi_\ell(2-\phi_\ell)/T},
\ena
where $\phi_\ell=v_{10}n_{\ell 0}$ is obtained from 
$p_{cx}^0\cong 0$ and 
  $\phi_g=v_{10}n_{g 0}$ from $\mu_{\rm cx}^0 
\cong T\ln (n_{g0}\lambda_1^3)$.

The quantity 
$\varphi$  in Eq.(2.2)  becomes  
\be 
\varphi=\frac{r\phi}{1-\phi} -
\ln(1-\phi)-\frac{2\epsilon }{T}w \phi,
\en
in terms of $\phi=v_0n_1$. 
Here two dimensionless parameters, the volume ratio and 
the potential ratio, are introduced as 
\be 
r=\frac{v_{02}}{v_{01}}, \quad w=\frac{w_{12}}{w_{11}},
\en 
which characterize the physical 
properties of the second component. 
If  $n(z)=n_1(z)$ is the density profile 
of the pure fluid across an interface, 
the density $n_2$ is expressed as in Eq.(2.9).  With the aid of Eqs.
(2.7) and (3.3) we rewrite $\varphi$ as 
\bea  
\varphi&=&\frac{1}{T}{\mu}_{\rm cx}^0 -\ln(n\lambda_1^3)+ D_{11}n''\nonumber\\
&& 
+ \frac{r-1}{1-\phi}\phi -  \frac{2\epsilon}{T}(w-1)\phi, 
\ena 
in terms of 
$\phi=v_{10}n(z)$. 
From  Eq.(2.9) the space-dependent 
molar fraction  $X(z)=n_2(z)/n_1(z)$ becomes  
\be 
X= {\tilde\zeta} \exp\bigg[\frac{1-r}{1-\phi}\phi
+ \frac{2\epsilon}{T}(w-1) \phi 
+D' n''\bigg],
\en   
where 
${\tilde \zeta} 
= \lambda_1^{3}e^{-{\mu}_{\rm cx}^0/T} \zeta.
= (m_2/m_1)^{3/2} e^{(\hat{\mu}_2-{\mu}_{\rm cx}^0)/T}$ and 
$D'=D_{12}-D_{11}$.  Notice that $X=$const. 
or ${\cal K}=1$  for $r=1$, $w=1$, 
and $D_{12}=D_{11}$, where  the two components 
have the same  physical properties.

\subsection{Two-phase coexistence}

\begin{figure}[htb]
\begin{center}
\includegraphics[scale=0.5]{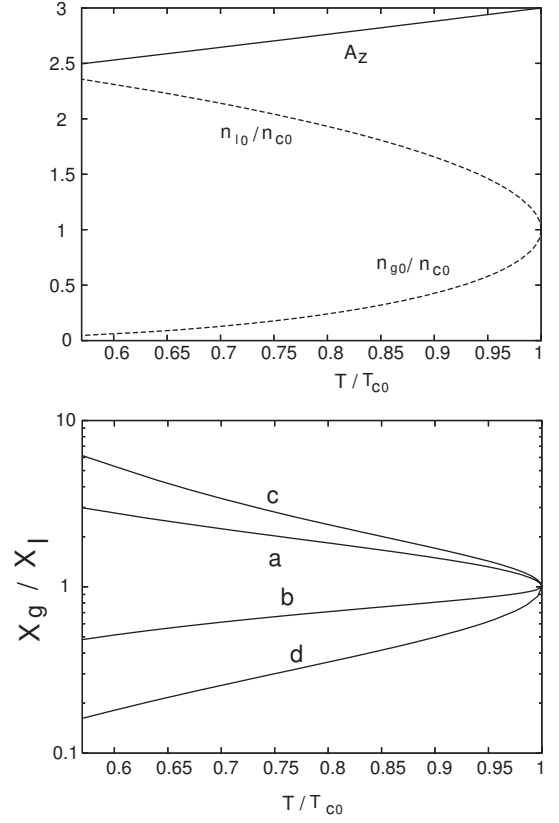}
\caption{Upper plate:  $A_z(T)$ in Eq.(3.13), 
$n_{\ell 0}/n_{c0}$, and  $n_{g 0}/n_{c0}$   
vs $T/T_{c0}$ in the van der Waals theory. 
Lower plate: ${\cal K}=X_g/X_\ell$ vs $T/T_{c0}$   
in  dilute mixtures on a semi-logarithmic scale, 
where  $(r,w)=$ 
$(0.8,0.8)$ for (a), 
$(0.8,1.0)$ for (b), 
$(1.5,1.0)$ for (c), and 
$(1.5,1.4)$ for (d).  Two parameters 
 $r$ and $w$ are defined in Eq.(3.9). 
}
\end{center}
\end{figure}

\begin{figure}[htbp]
\begin{center}
\includegraphics[scale=0.28]{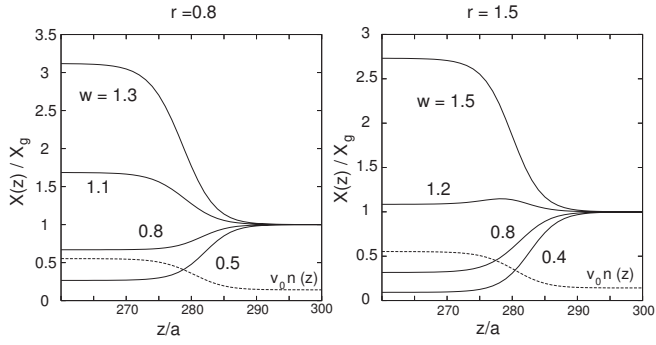}
\caption{	
Profiles of the 
 normalized molar fraction  $X(z)/X_g$ of the 
solute  at $T=0.9T_c$ 
for $r=0.8$ and $w=1.3, 1.1, 0.8$, and 0.5 (left) 
and for $r=1.5$ and $w=1.5, 1.2, 0.8$, and 0.4 (right).  
It tends to ${\cal K}^{-1}$ on the left hand side. 	
Shown also is the normalized density $v_0 n(z)$ 
of the pure fluid composed of 
the first component (broken line). Space is measured 
in units of the molecular size 
$a=v_{10}^{1/3}$. 
}
\end{center}
\end{figure}

From Eq.(3.11) the logarithm of the partition coefficient  
${\cal K}$ in Eq.(2.24) 
is expressed as 
\be 
\ln {\cal K}
= \frac{(r-1)\Delta \phi}{(1-\phi_\ell)(1-\phi_g)}
- \frac{2\epsilon}{T}(w-1) \Delta \phi,
\en   
where $\phi_\ell= v_{01} n_{\ell 0}$, 
 $\phi_g= v_{01} n_{g0}$,   and 
$\Delta\phi=\phi_\ell-\phi_g=v_{10}\Delta n$. 
In the lower plate of Fig.1, 
 $\cal K$ vs $T/T_{c0}$ is shown 
for typical four cases. 
Remarkably, the azeotropy (${\cal K}=1$) is attained 
in the dilute limit on the following line 
in the $r$-$w$ plane, 
\be 
{r-1}= A_z(T) ({w-1}), 
\en 
where the coefficient 
$A_z(T)$ is determined by the solvent 
properties only as  
\be 
A_z(T)= 2(1-\phi_\ell)(1-\phi_g)\epsilon/{T}.
\en 
See   the upper plate of Fig.1 for  $A_z(T)$ vs $T/T_{c0}$. 
Here $A_z \to 3$ as  $T \to T_{c0}$, 
while  for $\phi_g \ll 1$  we find $A_z \cong 2/\phi_\ell$ 
from  Eqs.(3.6) and (3.7) 
and  
\be 
\ln{\cal K} \cong 
[{r-1}- A_z ({w-1})]\phi_\ell^2 \epsilon/T.
\en
Using $k_H$ in Eq.(2.26), 
Sengers {\it et al}.\cite{Sengers1} examined 
$k_H/f_{0}$ where $f_0=\exp(\mu_0/T)$ is the fugacity 
of the pure fluid. In the van der Waals theory it is of the form, 
\be 
\ln \frac{k_H}{f_0} 
=\frac{3}{2} \ln \frac{m_1}{m_2}  
+ \frac{r-1}{1-\phi_\ell}\phi_\ell 
- \frac{2\epsilon}{T}(w-1)\phi_\ell.
\en

\begin{figure}[htbp]
\begin{center}
\includegraphics[scale=0.5]{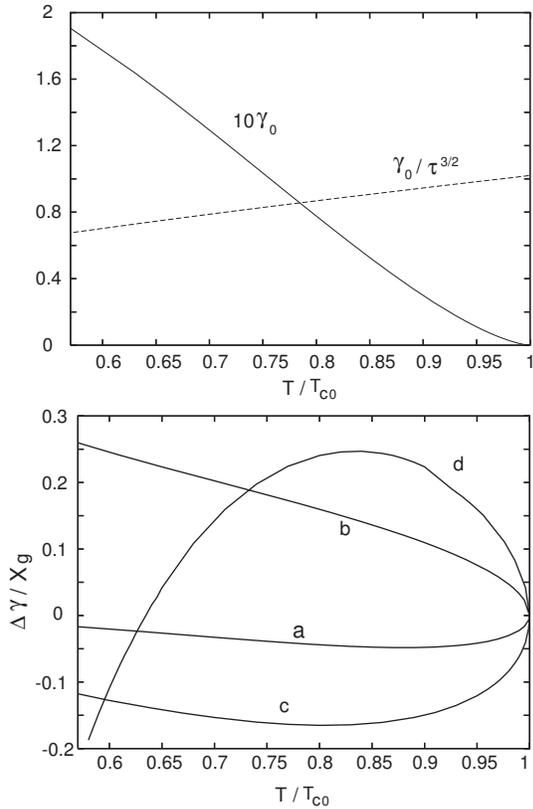}
\caption{Upper plate: 
Surface tension $\gamma_0$ multiplied by 10 
and $\gamma_0/\tau^{3/2}$ 
vs $T/T_{c0}$  for the pure fluid. 
Lower plate: $\Delta \gamma/X_g =-T\Gamma/X_g$ 
vs $T/T_{c0}$ in  dilute mixtures, where $(r,w)=$ 
$(0.8,0.8)$ for (a), 
$(0.8,1.0)$ for (b), 
$(1.5,1.0)$ for (c), and 
$(1.5,1.4)$ for (d).
The $\gamma_0$ and $\Delta\gamma$ 
are measured in units of $\epsilon/a^2$.
}
\end{center}
\end{figure}

\begin{figure}[htbp]
\begin{center}
\includegraphics[scale=0.35]{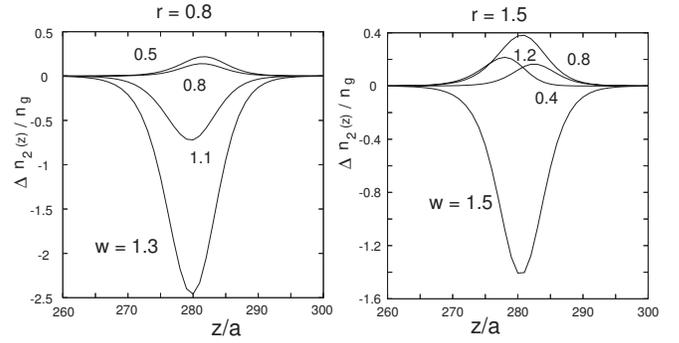}
\caption{ Excess solute density 
$\Delta n_2(z)$ in Eq.(3.17) divided by the molar fraction in the gas $X_g$ 
in units of $v_{10}^{-1}$.}
\end{center}
\end{figure}

\begin{figure}[htbp]
\begin{center}
\includegraphics[scale=0.5]{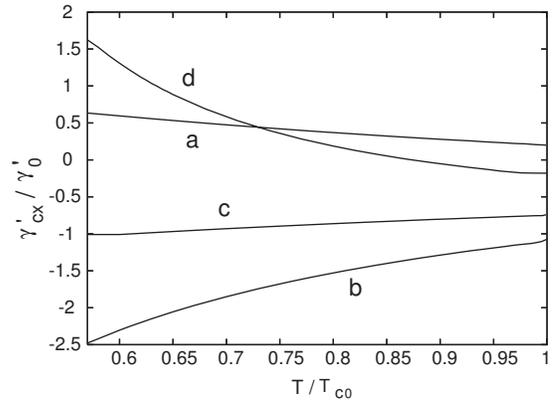}
\caption{
Ratio  of the temperature derivatives 
$\gamma'_{\rm cx}= (\p \gamma/\p T)_{{\rm cx},p}$ 
and $\gamma'_{0}= d \gamma_0/d T$  
vs $T/T_{c0}$ in  dilute mixtures, 
where $(r,w)=$ 
$(0.8,0.8)$ for (a), 
$(0.8,1.0)$ for (b), 
$(1.5,1.0)$ for (c), and 
$(1.5,1.4)$ for (d). It changes its sign for (d). 
}
\end{center}
\end{figure}

In Fig.2, we display profiles of 
the  molar fraction $X(z)$ 
divided by the molar fraction in the gas region $X_g$ 
at $T=0.9T_c$, where $A_{z}= 2.88$. 
In the left panel, we set 
$r=0.8$ and vary  $w$  as 
$1.3, 1.1, 0.8$, and 0.5. 
In the right panel, at $r=1.5$, we have 
$w=1.5, 1.2, 0.8$, and 0.4.  
Thus  
$X_\ell/X_g= {\cal K}^{-1}$ increases 
with decreasing $r$ and/or with increasing $w$.  
In our numerical analysis, 
we set  $D_{11}=D_{12}=10a^{5}$ 
with $a=v_{10}^{1/3}$. See the next subsection for 
justification of this choice of $D_{11}$. 
It is worth noting that Sahimi and Taylor calculated 
the density profiles of  two components 
around an interface \cite{Sahimi}.

\subsection{Surface tension}

In the upper plate of Fig.3, we show $\gamma_0$ and 
$\gamma_0/\tau^{3/2}$ vs $T/T_{c0}$ for the pure fluid, 
where we used the formula (A5) in Apenndix A  with 
$D_{11}=10a^{5}$. 
The relation $\gamma_0 \propto \tau^{3/2}$ 
nicely holds over a wide range of $T/T_{c0}$. 
Remarkably,  experimental data of the surface tension  of water 
can also be nicely  fitted  to the formula 
$\gamma_{\rm exp}= A_{exp}(1-T/T_{c0})^{3/2}$ 
over a  wide temperature range except  close to 
the criticality (in the range 
$1-T/T_{c0}\gs 0.1)$ \cite{Kie}. 
As in our previous work \cite{Kitamura}, 
we have determined  $D_{11}$  such that 
our numerical $\gamma$ and the experimental 
$\gamma_{\rm exp}$ for water reasonably  agree 
except close to the criticality. 
In fact,  at $T/T_c=0.675$, our $\gamma$ 
is   42.5 dyn$/$cm  if we  set   $D_{11}=10a^{5}$, $a=3{\rm \AA}$, 
 and $T_{c0}=647.1$K, 
while  the experimental value of water is 
44.6 dyn$/$cm.

In the lower  plate of Fig.3, we display the surface tension change 
 $\Delta\gamma=-T \Gamma$ 
divided by $X_g$ vs $T/T_{c0}$   for four sets of $(r,w)$. 
From Eq.(2.27)  $\Gamma$ is  the space integral of the excess solute density 
$\Delta n_2(z)$ expressed in terms of the 
density $n(z)$ of the reference pure fluid,  
\be 
\Delta n_2(z)= n_2(z)- n_{2\ell}- \frac{n_{2\ell}- n_{2g}}{n_{\ell 0}-n_{g0}}
 (n(z)- n_{\ell 0}).
\en 
In Fig.4, we plot $n_2(z)$ 
for $r=0.8$ (left) and $r=1.5$ (right) 
for various $w$. 
With increasing $w$,  $\Gamma$ becomes negative and 
its magnitude increases strongly. 
In Fig.5, we display 
the ratio $(dp_c/dX)/K_{\rm Kr}= 
(\p \gamma/\p T)_{{\rm cx},p}/(d\gamma/dT)$ 
calculated from Eq.(2.31) as a function of  
$T/T_{c0}$ for four sets of $(r,w)$. 
It even changes its sign from positive to negative 
with increasing $T$ for 
  $(r,w)=(1.5,1.4)$.

\subsection{Near-critical behavior}

The Landau expansion of $f_0$ with respect to 
$\psi=n-n_{c0}$ is given in Eq.(2.32). 
For  the van der Waals model  
the coefficients are given by   
\be 
A_0=\frac{27}{4} v_{10},
\quad B_0= \frac{243}{16} T_{c0}v_{10}^3.
\en  
In  the pure fluid,    
the liquid and gas densities 
are $n_{\ell 0}= n_{c0} + \psi_e$ 
and $n_{g 0}= n_{c0} - \psi_e$, where Eq.(2.49) gives  
\be 
\psi_e= 
2n_{c0}\tau^{1/2}. 
\en 
Here $\tau=1-T/T_{c0}$ is 
the reduced temperature (positive below the critical 
temperature). 
See the upper plate of Fig.1 for $n_{\ell 0}$ and $n_{g0}$.  
Then,   
\be 
\Delta n=2\psi_e,\quad 
\Delta v= 18v_{10}^2\psi_e, \quad  
\Delta s= 9v_{10}\psi_e.  
\en 
These differences are of order $\tau^{1/2}$. 
In particular, $\Delta s=6\tau^{1/2}$. 
The latter two relations are consistent with 
the Clausius-Clapeyron relation 
$\Delta s/\Delta v= (\p p/\p T)_{\rm cx}= 
1/2v_{10}$  along the 
coexistence curve.  In addition, 
the correlation length $\xi$ in Eq.(2.51) 
becomes $\xi= 0.86 a  \tau^{-1/2}$ 
in our numerical analysis with  $D_{11}=10a^5$.

\begin{figure}
\begin{center}
\includegraphics[scale=0.5]{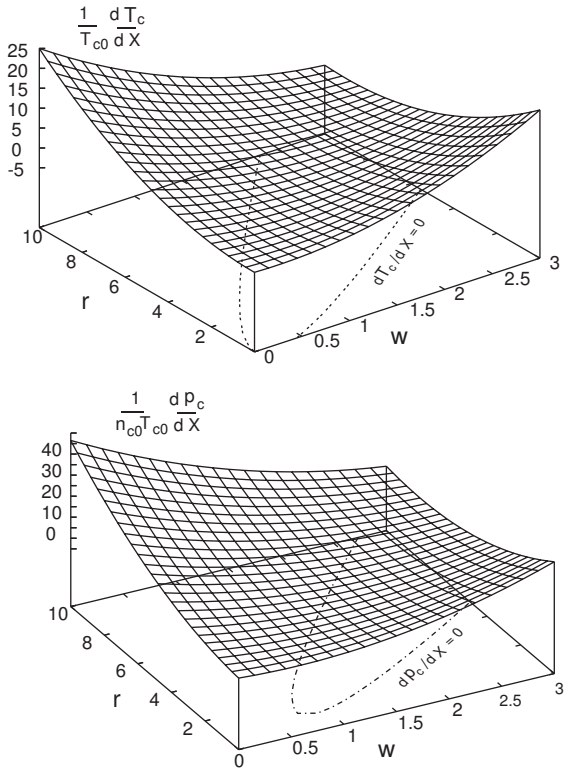}
\caption{$T_{c0}^{-1}dT_c/dX$ (upper plate) 
and $(n_{c0}T_{c0})^{-1}dp_c/dX$ (lower plate) 
as functions of $r$ and $w$ in Eq.(3.9) 
 near the solvent critical point.}
\end{center}
\end{figure}

\begin{figure}
\begin{center}
\includegraphics[scale=0.52]{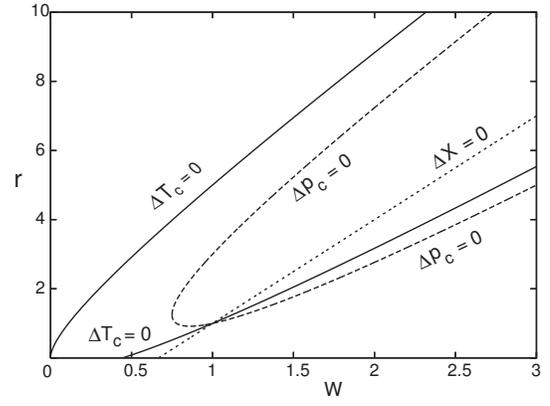}
\caption{Curves of 
$\Delta T_c=0$, $\Delta p_c=0$, and $\Delta X=0$ 
on the plane  of $(r,w)$ 
near the solvent critical point.  
}
\end{center}
\end{figure}

Using  $\varphi$ in Eq.(3.6) we perform the Taylor 
expansion $e^{-\varphi}
=C_0+ C_1\psi+C_2\psi^2/2+\cdots$ 
 as in Eq.(2.36). In terms of $r$ and $w$ 
in Eq.(3.9) the coefficients are expressed as    
\bea 
C_0&=& \frac{2}{3} \exp(-r/2+9w/4),\nonumber\\
C_1&=& \frac{3}{4} v_{10}(-3r-2+9w)C_0,\nonumber\\
C_2&=& \frac{81}{16} v_{10}^2[(r-3w)^2-4w]C_0.
\ena 
The critical solute  density 
 and   concentration are 
\be 
n_{2c}= C_0\zeta, \quad  
X_c= 3v_{10}C_0\zeta.
\en 
The solute  density difference in Eq.(2.18) 
and the composition difference in Eq.(2.19) 
are expressed  as 
 \bea 
&&{\Delta n_2}
=(3r-9w+2)X_cn_{c0} \tau^{1/2},\nonumber\\ 
&& 
\Delta X 
= 3(r-3w+2) X_c \tau^{1/2}.
\ena 
The Krichevskii parameter in Eq.(2.47) is given by 
\be 
K_{\rm Kr}/n_{c0}T_{c0} = \frac{3}{4}(r-3w+2), 
\en 
which was already derived by 
Petsche and Debenedetti \cite{De}. 
See Table 1 for experimental values of the above quantity. 
In accord with these results, ${\cal K}$ behaves as 
\be 
{\cal K}=1+ 3(r-3w+2)\tau^{1/2} +\cdots. 
\en 
while $\ln (k_H/f_0)= {\rm const.}+ 
3(r-3w+2)\tau^{1/2}/2+\cdots$ 
from Eq.(3.16).  
Sengers {\it et al.} \cite{Sengers1} found that 
data of $T\ln (k_H/f_0)$ can well be fitted to 
the form $C +B (n_\ell -n_{c0})$ near the critical point 
for a number of solutes in H$_2$O.

From Eqs.(2.43) and (2.44) 
the derivatives 
$dT_c/dX$ and 
$dp_c/dX$ along the critical line are 
written as  
\bea 
&&\hspace{-1cm} \frac{1}{T_{c0}} \frac{d T_c}{dX} =  
 \frac{1}{4}(r-3w)^2-w,\\
&& \hspace{-1cm} \frac{1}{p_{c0}} \frac{d p_c}{dX} =     
 (r-3w)(r-3w+2) -4(w-1) .
\ena 
In Fig. 6, we show 
$d T_c/dX$ and  $d p_c/dX$ in the $r$-$w$ plane. 
In Fig.7, we show the curves of 
$d T_c/dX=0$, $d p_c/dX=0$,  
and the azeotropic line $\Delta X=0$. 
Thus, $\Delta X$,  $dT_c/dX$, and $d p_c/dX$
 can be both positive and negative depending on $r$ and $w$.

From Eqs.(2.58) and (3.26) the surface adsorption 
$\Gamma$ is written as   
\be 
\Gamma= \frac{1}{4} [(r-3w)^2-4w]X_c\frac{d\gamma_0}{dT}. 
\en 
From Eq.(2.60) 
we calculate the temperature-derivative of $\gamma$ 
on the coexistence surface, 
\be 
\ppp{\gamma}{ T}{{\rm cx},p}=
\bigg[ \frac{r-3w}{2}- \frac{2(w-1)}{r-3w+2} \bigg] 
\frac{d\gamma_0}{dT}  .
\en 
Thus the above derivative 
can be both negative and positive 
and can even diverge to $\pm \infty$ 
on the azeotropic line  $r-3w+2= 0$.

\section{Summary} 
In summary,  a Ginzburg-Landau theory has been presented 
for  dilute binary mixtures,  
 where  the solute-solvent interaction is relevant but 
the solute-solute interaction is negligible. 
A parameter $\zeta$ 
proportional to the solute fugacity 
has been introduced in Eq.(2.8). 
Up to first order in $\zeta$, 
all the physical quantities of binary mixtures 
can easily be calculated in terms of the properties of 
the one-component fluid and the solute-solvent interaction 
parameters. In more detail, 
our main results are as follows\\
(i)  The coexistence surface  has been  given by 
Eqs.(2.16) and (2.23)  or by Eq.(2.22).
 Henry's constants  have been introduced in 
Eqs.(2.24)-(2.26). \\
(ii) The Gibbs formula for the 
surface tension change $\Delta\gamma$ in Eq.(2.27) 
has been derived in Appendix A. 
The surface tension derivative $(\p \gamma/\p T)_{p{\rm cx}}$ 
with respect to $T$ at fixed $p$ 
has been obtained  in Eq.(2.31). Interestingly,  
it consists of two terms both being independent of $\zeta$. 
\\ 
(iii) The critical 
temperature shift $\Delta T_c $ is given 
in Eq.(2.41)  and 
the critical pressure shift $\Delta p_c $ in Eq.(2.42).\\
(iv) The Krichevskii parameter $K_{\rm Kr}$ 
has been given in Eqs.(2.47) and (2.48).  
The normalized parameter  $K_{\rm Kr}X_c/n_{c0}T_{c0}$ 
 represents the size of the critical 
concentration fluctuations as in Eq.(2.49), 
leading to Eq.(2.52).\\
(v) The surface adsorption  
$\Gamma$ has been realted 
to $\Delta T_c$ as in Eq.(2,58) 
and $(\p\gamma/\p T)_{{\rm cx},p}$ to $\Delta p_c$ 
as in Eq.(2,66) near the criticality. 
The solute effect on the near-critical 
surface tension  is simply to shift  the 
critical temperature $T_{c0}$ by $\Delta T_{c}$ as in Eq.(2.59).
\\ 
(vi) Experimental 
data  of $dT_c/dX$,  $dp_c/dX$, and $K_{\rm Kr}$ 
have been given 
 in scaled forms in Table 1, which  shows  
that they can be both 
positive and negative.\\  
(vii) The van der Waals model of 
binary mixtures 
has given simple expressions for all 
the theoretical expressions in Section II, as illustrated 
in the figures. 
The solute-solvent interaction is described in terms of 
the size ratio $r$ and the potential ratio $w$ in Eq.(3.9). 
\\
(viii) The profiles of the 
solute density  and its  excess 
near an  interface 
have been numerically calculated as in  Fig. 2 and 4. 
The negative  adsorption 
becomes marked for large $w$. 
\\ 
(ix) The near-critical behavior 
in the van der Waals model is very simple 
in the mean-field theory. 
In terms of $r$ and $w$ we have 
calculated $\cal K$, 
$K_{\rm Kr}$, $dT_c/dX$, $dp_c/dX$, 
$\Gamma$, and $(\p \gamma/\p T)_{{\rm cx},p}$. 
In the $r$-$w$ plane , we have plotted    $dT_c/dX$ and  $dp_c/dX$ 
in Fig.6 and the curves 
of  $dT_c/dX=0$,   $dp_c/dX=0$, 
and ${\cal K}=1$  in Fig. 7. 
\\

Finally, we propose  measurements of 
the surface tension as a function of the temperature 
in the isobaric condition  
for various solutes in water or in CO$_2$.   
 From our theory, the  derivative $(\p \gamma/\p T)_{{\rm cx}.p}$ 
becomes independent of the solute density 
in the dilute limit and  
can be both negative and positive, being delicately dependent 
on the size ratio $r$ and the potential ratio $w$ 
in the van der Waals theory 
as in Eq.(3.29). It is a relevant parameter 
determining the Marangoni flow around 
a bubble moving in heat flow in  
binary mixtures, as will  be reported shortly.

\vspace{2mm} 
{\bf Appendix A: Calculation of surface tension}\\
\setcounter{equation}{0}
\renewcommand{\theequation}{A\arabic{equation}}

Here  the surface tension $\gamma$ of 
binary mixtures is examined from  Eq.(2.1). 
The grand potential density of mixtures 
is given  by   
\be 
\omega  =   f -\sum_i \hat{\mu}_i n_i 
 + \frac{T}{2}\sum_{i,j} D_{ij} n_i' n_j' ,
\en
where $n_i'=dn_i/dz$ and 
$\hat{\mu}_i$ take the values in two-phase coexistence. 
All the quantities change  along the $z$ axis.   
The space integral of $\omega$ gives the grand potential 
$\Omega$ in Eq.(2.10). Then $\omega$ tends  to 
$-p_{\rm cx}$  far from the interface 
$z\to \pm \infty$ and the surface tension is expressed as 
\be 
\gamma=\int dz[\omega(z)+p_{\rm cx}].
\en 
Differentiation of $\omega(z)$ in Eq.(A1) with respect to $z$ 
yields $d\omega /dz=2T\sum_{ij}D_{ij} n_i'n_j''$ 
from Eq.(2.6), where $n_j''= d^2n_j/dz^2$.  Therefore,  
\bea 
\omega &=& T\sum_{ij}D_{ij} n_i'n_j'-p_{\rm cx}\nonumber\\
&=& 2(f -\sum_i \hat{\mu}_i n_i) +p_{\rm cx}.
\ena 
Then $\gamma$ in Eq.(A2) may also be expressed as 
 $\gamma=\int dzT \sum_{ij}D_{ij} n_i'n_j'
=2\int dz (f -\sum_i \hat{\mu}_i n_i +p_{\rm cx})$.

Next  $\gamma$ is expanded 
with respect to $\zeta$ in  the dilute case. 
As in the derivation of Eq.(2.11),  
 elimination of  
$\hat{\mu}_2$ in Eq.(A1) gives 
\be 
{\gamma}  = \int dz \bigg [ f_0(n_1) -
{\hat{\mu_1}} n_1 
+  p_{\rm cx} + 
\frac{T}{2}D_{11} n_1'^2
-  Tn_2 \bigg], 
\en
where $\hat{\mu}_1= \mu_{\rm cx}^0+ 
\delta\mu_{1{\rm cx}}$, 
 $n_2$ is given by the second line of 
Eq.(2.9),  and the integrand vanishes  as $z\to \pm \infty$.  
Let $n=n(z)$ be the density of the reference pure fluid or
$n(z)=\lim_{\zeta\rightarrow 0} n_1(z).$ Then 
$n(z) \to n_{\ell 0}$ 
($n_{g 0}$) as $z \to -\infty$ ($\infty)$ 
and   we have 
the interface equation (2.7). 
As $\zeta\to 0$ the surface tension of  the pure  
fluid is obtained  as 
\be 
{\gamma_0}= 
\int dz  \bigg [ f_0(n)-\mu_{\rm cx}^0 n+ 
p_{\rm cx}^0 + \frac{T}{2}{D_{11}}n'^2\bigg].
\en  
From Eqs.(A4) and (A5) 
the surface tension change 
$\Delta\gamma=\gamma-\gamma_0$ for small $\zeta$ 
is expanded with respect to the 
deviation $\delta n_1(z)=n_1(z)-n(z)$ as 
\bea 
\Delta\gamma&=&
\int dz  \bigg
[ (f_0'(n)-\mu_{\rm cx}^0-TD_{11}n'')\delta n_1 \nonumber\\
&&
-\delta \mu_{1\rm cx} n+ 
\delta p_{\rm cx} -Tn_2\bigg]+\cdots,
\ena 
to first order in $\zeta$. Here the first term in the brackets 
vanishes from Eq.(2.7).  Further use of   Eqs.(2.16) and (2.17) 
yields  the Gibbs relation in Eq.(2.27).

\vspace{2mm} 
{\bf Appendix B: Correlation-function expressions}\\
\setcounter{equation}{0}
\renewcommand{\theequation}{B\arabic{equation}}

We examine the correlation-function expressions  
for thermodynamic derivatives such as 
$K_{\rm Kr}$ in Eq,(2.47) 
and $(\p X/\p \Delta)_{pT}$ in 
Eq.(2.52) in the framework in the book 
of  the present author \cite{Onukibook}. 
Equivalent relations for  $K_{\rm Kr}$  were 
already used in the literature  \cite{OC,De,Shock}.

The microscopic particle densities are written as  
\be
\hat{n}_j({\bi r})=\sum_{\ell\in j}
\delta({\bi r}-{\bi r}_\ell),
\en  
where the summation is over the particles of the species $j 
(=1,2)$ at position ${\bi r}_\ell$. 
Then $n_j=\av{\hat{n}_j}$, where $\av{\cdots}$ 
denotes  the equilibrium average. The pair correlation functions 
are written as 
\be 
\av{\delta \hat{n}_i({\bi r})
\delta \hat{n}_j({\bi 0})}
=n_i \delta_{ij}\delta({\bi r}) +n_in_jg_{ij}(r),
\en 
where $\delta\hat{n}_i({\bi r})= 
\hat{n}_i({\bi r})-n_i$ $(i=1,2)$ are the density deviations 
and 
 $g_{ij}(r)$ ($i, j=1,2)$  are 
 the radial distribution functions tending to zero 
for large separtion $r$. 
It is convenient to introduce 
the concentration variable ${\hat X}({\bi r})$ and 
the number density 
variable $\hat{n}({\bi r})$ by 
\be 
{\hat X}=X+  \frac{n_1}{n^2}\hat{n_2}-\frac{n_2}{n^2}\hat{n_1}, 
\quad 
{\hat n}=\hat{n_1}+\hat{n_2}, 
\en 
where $\av{\hat{n}}= 
n=n_1+n_2$ and $\av{\hat{X}}=X=n_2/n$.  
We define a    fluctuation 
variance for any 
space-dependent variables $\hat{A}({\bi r})$ 
and  $\hat{B}({\bi r})$ by  
\be 
\av{\hat{A}:\hat{B}}= \int d{\bi r}
\av{(\hat{A}({\bi r})-\av{\hat{A}}) 
(\hat{B}({\bi r})-\av{\hat{B}}) }.
\en
The variances among $\hat n$ and $\hat X$ 
may be expressed in terms of 
 the thermodynamic derivatives, 
\bea 
&&\av{\hat{n}:\hat{n}}= nT \ppp{n}{p}{T\Delta},
\quad 
\av{\hat{X}:\hat{X}}= \frac{T}{n}\ppp{X}{\Delta}{pT}, 
 \nonumber\\ 
&&\av{\hat{n}:\hat{X}}= nT \ppp{X}{p}{T\Delta}= \frac{T}{n}
\ppp{n}{\Delta}{pT},
\ena 
where $n$ and $X$  are 
treated as functions of the field variables $T$, $p$, and 
$\Delta=\mu_2-\mu_1$ in the derivatives. 
These variances are linear combinations 
of the variances among the densities, which 
are written as 
\be 
I_{ij}\equiv 
\av{\hat{n}_i:\hat{n}_j}=n_i \delta_{ij} +  n_in_j 
\int d{\bi r}g_{ij}(r), 
\en 
from Eq.(B2).  
On the other hand, the compressibility at constant $X$ 
is written as 
\be 
K_{TX}= \frac{1}{n}\ppp{n}{p}{TX}
= \frac{1}{n^2T}\bigg[\av{\hat{n}:\hat{n}}-
\frac{\av{\hat{n}:\hat{X}}^2}{\av{\hat{X}:\hat{X}}}\bigg], 
\en  
Near the mixture criticality,  the ratio 
$K_{TX}/ K_{T\Delta}$ behaves as $X/
\av{\hat{X}:\hat{X}} 
\cong nX/[(\p X/\p \Delta)_{Tp}T]$ (see Eq.(2.52)). 
All the variances in Eqs.(B5) and (B6) diverge strongly at the 
mixture criticality except for special cases such as 
the critical azeotropy. 
In the low density limit $X\to 0$ 
under Eq.(2.44),  Eqs.(B3) and (B6) give  
\be 
\av{\hat{X}:\hat{X}}\cong \av{\hat{n}_2:\hat{n}_2}/n^2 
\cong X/n.
\en 
We also need to  assume $\av{\hat{X}:\hat{n}}
\propto  X$  for the existence of the 
Krichevskii parameter (see Eqs.(B9) and (B10)).

We next examine the thermodynamic derivative $
({\p p}/{\p X})_{nT}=  
- (\p{n}/\p{X})_{pT}/{nK_{TX}}$. 
Its  correlation-function expression reads  
\be 
\ppp{p}{X}{nT}= \frac{-nT\av{\hat{n}:\hat{X}}}{
\av{\hat{n}:\hat{n}}\av{\hat{X}:\hat{X}}-
\av{\hat{n}:\hat{X}}^2 }.
\en
In the low density limit we use  Eq.(B8) 
and replace the denominator of 
Eq.(B9) by $\av{\hat{n}:\hat{n}}X/n$ to  find 
\bea
\lim_{X\to 0}\frac{1}{nT}\ppp{p}{X}{nT} &=&  -\lim_{X\to 0} 
\frac{n\av{\hat{n}:\hat{X}}}{X\av{\hat{n}:\hat{n}}}\nonumber\\
&=& 1-n_1C_{12}^\infty, 
\ena
where the second line follows from Eq.(B3). We define 
\be
C_{12}^\infty = 
\lim_{n_2\to 0} 
\av{\hat{n}_2:\hat{n}_1}/n_2\av{\hat{n}_1:\hat{n}_1},  
\en 
which coincides  with 
the space-integral of the 
direct correlation function $C_{12}(r)$  in the dilute 
limit \cite{De,OC,Shock}. Here we define 
$C_{ij}(r)$ in dimensionless forms \cite{Onukibook}. 
Thus the  Krichevskii parameter $K_{\rm Kr}$  in  Eq.(2.47) is the 
 value  of $n_1T(1-n_1C_{12}^\infty)$ 
at the solvent criticality. This expression  has been 
used to  estimate  $K_{\rm Kr}$ 
for given  molecular interaction parameters 
\cite{OC,De,Shock}. 
From Eqs.(B10) and (B11)  
the singular parts of 
$\hat{X}$ and $\hat{n}_2$ 
are 
\bea 
(\hat{X})_{\rm sing}&=& 
(C_{12}^\infty-1/n_1)X\delta{\hat n}_1,\nonumber\\ 
(\hat{n}_2)_{\rm sing}&=& 
C_{12}^\infty n_2\delta{\hat n}_1,
\ena  
near the mixture criticality. 
Here we have calculated {\it projected} 
parts of $\hat X$ and ${\hat n}_2$ 
onto  the critical fluctuation 
$\delta{\hat n}_1=\hat{n}_1-n_1$. 
Equation (2.52) 
is then obtained with the aid of Eq.(B8).

{\bf Acknowledgments}\\
I would like to thank Dr. J.M.H. Levelt Sengers 
for informative correspondence on Henry's law near the criticality. 
This work was supported by KAKENHI (Grant-in-Aid for Scientific Research) on Priority Area gSoft Matter Physicsh from the Ministry of Education, Culture, Sports, Science and Technology of Japan.


\begin{references}
\bibitem{Maran} We consider  droplet 
motion in heat flow. For slow motions 
the pressure deviation is homogeneous 
around  a droplet,  while the  temperature deviation  
and   the chemical potential deviations 
at the interface are  
on the coexistence surface. 
The surface tension deviation is then 
$(\p \gamma/\p T)_{{\rm cx},p}\delta T$.
\end{references}
\end{document}